
\magnification=\magstep0
\hsize=13.5 cm               
\vsize=19.0 cm               
\baselineskip=12 pt plus 1 pt minus 1 pt  
\parindent=0.5 cm  
\hoffset=1.3 cm      
\voffset=2.5 cm      
\font\twelvebf=cmbx10 at 12truept 
\font\twelverm=cmr10 at 12truept 
\overfullrule=0pt
\input psfig.tex
%
\newtoks\leftheadline \leftheadline={\hfill {\eightit Authors' name}
\hfill}
\newtoks\rightheadline \rightheadline={\hfill {\eightit the running title}
 \hfill}
\newtoks\firstheadline \firstheadline={
\hfill}
\def\makeheadline{\vbox to 0pt{\vskip -22.5pt
\line{\vbox to 8.5 pt{}\ifnum\pageno=1\the\firstheadline\else%
\ifodd\pageno\the\rightheadline\else%
\the\leftheadline\fi\fi}\vss}\nointerlineskip}
%
\font\eightrm=cmr8  \font\eighti=cmmi8  \font\eightsy=cmsy8
\font\eightbf=cmbx8 \font\eighttt=cmtt8 \font\eightit=cmti8
\font\eightsl=cmsl8
\font\sixrm=cmr6    \font\sixi=cmmi6    \font\sixsy=cmsy6
\font\sixbf=cmbx6
%
\def\eightpoint{\def\rm{\fam0\eightrm}
\textfont0=\eightrm \scriptfont0=\sixrm \scriptscriptfont0=\fiverm
\textfont1=\eighti  \scriptfont1=\sixi  \scriptscriptfont1=\fivei
\textfont2=\eightsy \scriptfont2=\sixsy \scriptscriptfont2=\fivesy
\textfont3=\tenex   \scriptfont3=\tenex \scriptscriptfont3=\tenex
\textfont\itfam=\eightit  \def\it{\fam\itfam\eightit}%
\textfont\slfam=\eightsl  \def\sl{\fam\slfam\eightsl}%
\textfont\ttfam=\eighttt  \def\tt{\fam\ttfam\eighttt}%
\textfont\bffam=\eightbf  \scriptfont\bffam=\sixbf
\scriptscriptfont\bffam=\fivebf \def\bf{\fam\bffam\eightbf}%
\normalbaselineskip=10pt plus 0.1 pt minus 0.1 pt
\normalbaselines
\abovedisplayskip=10pt plus 2.4pt minus 7pt
\belowdisplayskip=10pt plus 2.4pt minus 7pt
\belowdisplayshortskip=5.6pt plus 2.4pt minus 3.2pt \rm}
%
%
\def\leftdisplay#1\eqno#2$${\line{\indent\indent\indent%
$\displaystyle{#1}$\hfil #2}$$}
\everydisplay{\leftdisplay}
%
\def\frac#1#2{{#1\over#2}}


%
\input psfig.tex
%
\def\pmb#1{\setbox0=\hbox{$#1$}\kern-0.015em\copy0\kern-\wd0%
\kern0.03em\copy0\kern-\wd0\kern-0.015em\raise0.03em\box0}
%
\def\teff{T$_{\rm eff}$}
\def\logg{$\log g$}
\def\met{[M/H]}

\def\yv{{\bf y}}

\def\xv{{\bf x}}

\def\qv{{\bf q}}
\def\wv{{\bf w}}

\def\sig68{$\sigma_{68}$}
\def\sigrms{$\sigma_{RMS}$}

%
%
%
\pageno=1
\vglue -28 pt
\leftline{\eightrm To appear in  {\eightit Automated Data Analysis in Astronomy},
	  R.\ Gupta, H.P.\ Singh, C.A.L.\ Bailer-Jones (eds.),} 
\leftline{\eightrm Narosa Publishing House, New Delhi, India, 2001}
\vglue 68 pt  
%
\leftline{\twelvebf  An introduction to artificial neural networks}
%
\smallskip
\vskip 46 pt  
\leftline{\twelverm Coryn A.L.\ Bailer-Jones} 
\vskip 4 pt
\leftline{\eightit Max-Planck-Institut f\"ur Astronomie, K\"onigstuhl 17, 69117 Heidelberg, Germany
}
\leftline{\eightit email: calj@mpia-hd.mpg.de}
%
%
\vskip 0.5 cm
\leftline{\twelverm Ranjan Gupta} 
\vskip 4 pt
\leftline{\eightit IUCAA, Post Bag 4, Ganeshkhind, Pune-411007, India}
\leftline{\eightit email: rag@iucaa.ernet.in}
\vskip 0.5 cm
\leftline{\twelverm Harinder P.\ Singh}
\vskip 4 pt
\leftline{\eightit Department of Physics, Sri Venkateswara College, Benito Juarez Road, Dhaula Kuan}
\leftline{\eightit New Delhi-110021, India}
\leftline{\eightit email: hps@ttdsvc.ernet.in}

\vskip 20 pt 
%
%
\leftheadline={\hfill {\eightit Bailer-Jones, Gupta \& Singh} \hfill}
\rightheadline={\hfill {\eightit An introduction to artificial neural networks
}  \hfill}

%
{\parindent=0cm\leftskip=1.5 cm

{\bf Abstract.}
\noindent
Artificial neural networks are algorithms which have been developed to
tackle a range of computational problems. These range from modelling
brain function to making predictions of time-dependent phenomena to
solving hard (NP-complete) problems.  In this introduction we describe
a single, yet very important, type of network known as a feedforward
network. This network is a mathematical model which can be trained to
learn an arbitrarily complex relationship between a data and a
parameter domain, so can be used to solve interpolation and
classification problems. We discuss the structure, training and
interpretation of these networks, and their implementation, taking the
classification of stellar spectra as an example.

}                                 

%
%
%
\vskip 20 pt
\centerline{\bf 1.\ Biological neural networks}
\bigskip
\noindent
Artificial neural networks were originally introduced as very
simplified models of brain function, so it is initially instructive to
consider the broad analogies between artificial neural networks and
biological ones. Of course, the latter are considerably more complex,
and many artificial neural network models today bear very little
resemblance to biological ones. For a history of the development of
artificial neural networks, see, for example, the introduction of
Hertz, Krogh and Palmer (1991).

The human brain consists of billions of interconnected neurons. These
are cells which have specialized membranes which allow the
transmission of signals to neighbouring neurons. Fig.~1 shows the
structure of a neuron.  A single {\it axon} extends from the basic
cell body.  This is the output from the neuron, and typically divides
into many sub-branches before terminating at a {\it
synapse}. Electrical pulses are transmitted along the axon to the
synapses by the transfer of Na$^{+}$ ions. The arrival of a voltage
pulse stimulates the release of neurotransmitting chemicals across the
synaptic cleft towards the postsynaptic cell, which is the receiving
part of the next neuron.  This postsynaptic cell passes the signal via
the {\it dendrite} to the main part of the neuron body.  The inputs
from the different dendrites are then combined to produce an output
signal which is passed along the axon, and so the process
continues. However, a signal is only produced in the axon if there are
enough inputs of sufficient strength to overcome some threshold value,
and then the output is some nonlinear function of the input stimuli.

\noindent
\midinsert
\vskip 8.3cm
\includegraphics{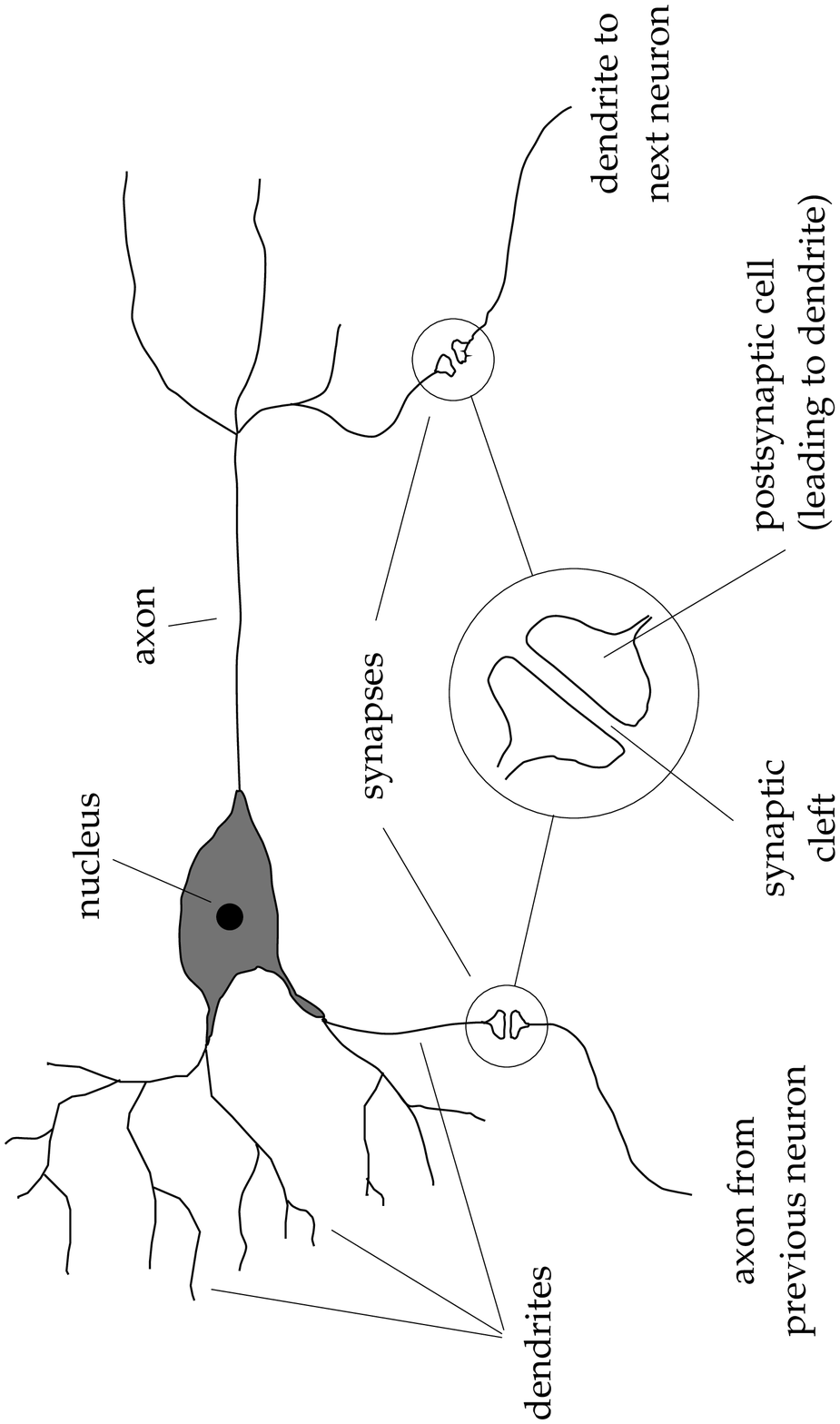}
\vskip 0.3cm
{\eightpoint   
\noindent
{\bf Figure 1.} Structure of a single neuron in the brain.
Information is passed into the neuron from other neurons via the
dendrites to the left. Given appropriate input stimuli, the cell sends
an output signal along the axon to the synapses, where the signal is
transmitted to other neurons.  }
\endinsert

The human brain consists of about $10^{11}$ neurons, and each neuron
has between a few and a few thousand synapses on its dendrites, giving
a total of about $10^{14}$ synapses (connections) in the brain. The
`strength' of the synaptic connection between neurons can be
chemically altered by the brain in response to favourable and
unfavourable stimuli, in such a way as to adapt the organism to
function optimally within its environment. The synapses are therefore
believed to be the key to learning in biological systems.

\vskip 12 pt
\centerline{\bf 2.\ Artificial neural networks}
\bigskip
\noindent
``Artificial neural networks'' is a relatively loose term referring to
mathematical models which have some kind of distributed architecture,
that is, consist of processing nodes (analogous to neurons) with
multiple connections (analogous to dendrites and axons). These
connections generally have adaptable parameters which modify the
signals which pass along them. There are numerous types of artificial
neural networks for addressing many different types of problems, such
as modelling memory, performing pattern recognition, and predicting
the evolution of dynamical systems. Most networks therefore perform
some kind of data modelling, and they may be split into two broad
classes: {\it supervised} and {\it unsupervised}. The former refers to
networks which attempt to learn the relationship between a data and a
parameter domain, while the latter refers to networks used to find
``natural'' groupings with a data set independently of external
constraints. What they have in common is the idea of learning about a
problem through relationships intrinsically present in data, rather
that through a set of predetermined rules.  An introduction to
several types of neural networks (but by no means all) is given by
Hertz, Krogh and Palmer (1991). A less mathematical overview is
provided by Beale \& Jackson (1990).

In this article we introduce what is one of the most important types
of supervised neural networks, called a {\it feedforward multilayer
perceptron}. The term {\it perceptron} is historical, and refers to
the function performed by the nodes. {\it Feedforward} means that
there is a definite input and output, and a flow of data in one
direction. This is in contrast to {\it recurrent neural networks} in
which data flows in a loop: these are important when time plays a
relevant role in the problem (see, e.g., Bailer-Jones 1998a). A
comprehensive treatment of feedforward networks is provided by Bishop
(1996), where more details on many of the themes discussed below can
be found.

\noindent
\midinsert
\vskip 9.5cm
\includegraphics{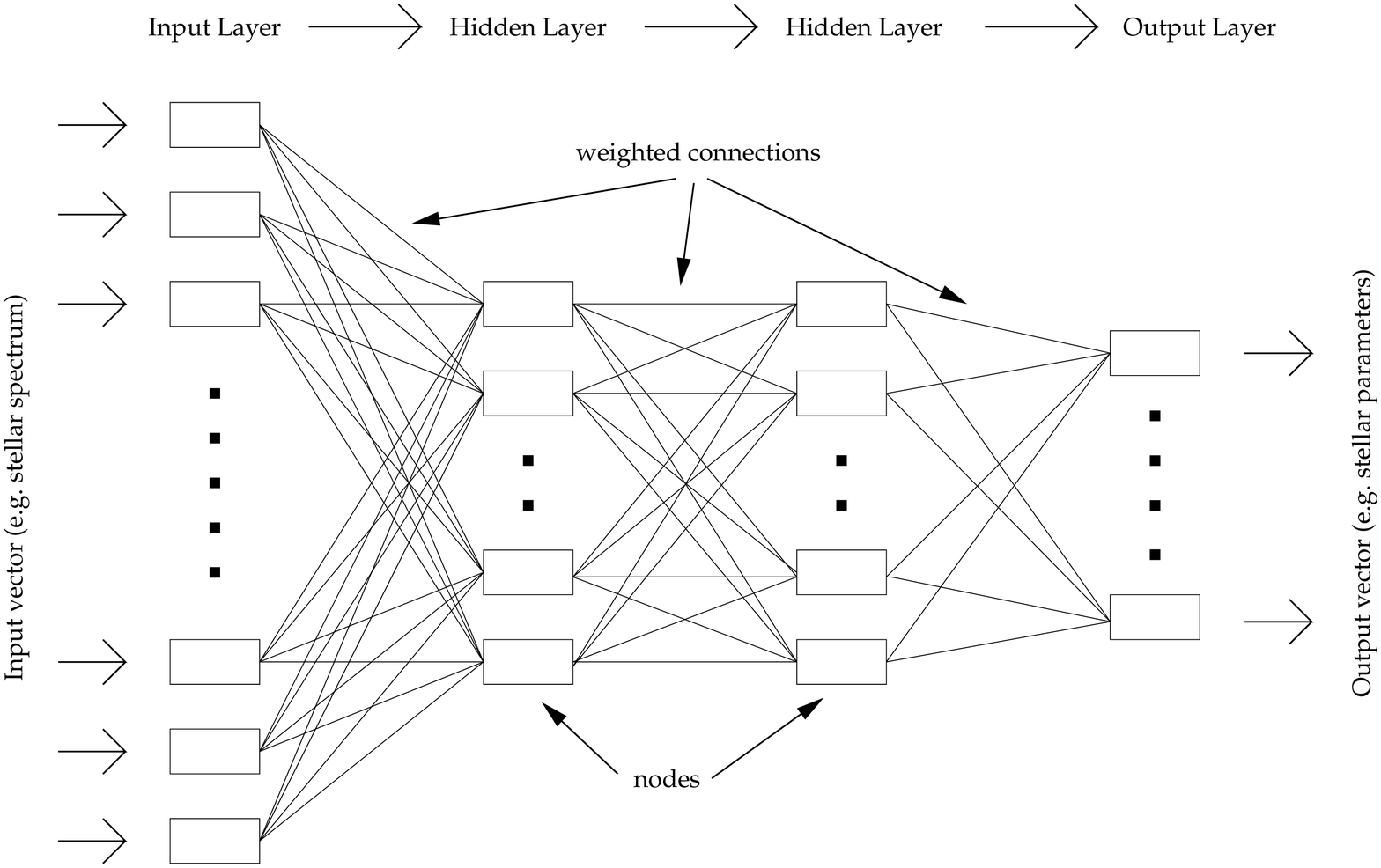}
\vskip 0.3cm
{\eightpoint   
\noindent
{\bf Figure 2.} Feedforward artificial neural network architecture.
This particular example shows two hidden layers. Not explicitly
shown are the bias nodes: each layer has an extra node that holds a
constant value (and has no inputs) and provides an offset to the
subsequent layers.  These are necessary for the network to model
functions properly.}
\endinsert

Fig.~2 shows a feedforward neural network with several inputs, two
hidden layers, and several outputs.  Each node in the input layer
holds a value, $x_i$. In our example application, the input vector,
$(x_1,x_2,\ldots,x_i,\ldots)$, is a stellar spectrum, and the output
vector, $(y_1,y_2,\ldots,y_l,\ldots)$, is a set of stellar parameters,
e.g.\ \teff, \logg\ and
\met. Each of the input nodes connects to every node in the next layer
of nodes, the first `hidden' layer, and each of these connections has
a weight, $w_{i,j}$, associated with it.  A node in the hidden layer
forms a weighted sum of its inputs, and passes this through a {\it
nonlinear transfer function}, such that the output from the $j^{th}$
hidden node is

$$ p_j = \tanh \left( \sum_i w_{i,j} x_i \right) \ \ . \eqno(1) $$

\noindent
These values are passed to a second hidden layer which performs
a similar processing, the output from that layer being the vector $\qv$

$$ q_k = \tanh \left( \sum_j w_{j,k} p_j \right) \ \ . \eqno(2) $$

\noindent
The output layer then performs a simple sum of its inputs, so that the
network output, $\yv$, is

$$ y_l =  \sum_k w_{k,l} q_k \ \ . \eqno(3) $$

\noindent
The tanh function in the hidden layers provides the nonlinear
capability of the network.  Other nonlinear functions are possible;
the sigmoidal function ($1/(1-\exp[-\sum wx])$) is often used. Both
functions map an infinite possible input range onto a finite output
range, $-1$ to $+1$ in the case of tanh. This imitates the transfer
function of neurons.

\vskip 30 pt

\centerline{\bf 3.\ Network training}
\bigskip
\noindent
The weights, $\wv$, appearing in equations 1--3 are the free parameters of the
network. Clearly, in order for the network to yield appropriate
outputs for given inputs, the weights must be set to suitable
values. This is done by {\it training} the network on a set of input
vectors for which the ideal outputs (or {\it targets}) are already
known. This is {\it supervised learning}. The idea is that the network
will encapsulate the relationship between, in our example, a wide
range of types of stellar spectra and their associated physical
parameters. The network does not come up with a new classification
system, but rather will {\it interpolate} the training data to give a
{\it generalization} of the relationship between spectra
and their classifications. We can then use this trained network to
obtain reliable classifications for unclassified spectra.

Supervised learning proceeds by minimizing an error function with
respect to all of the network weights. The error function typically
used is the sum-of-squares error, which for a single input vector,
$n$, is

$$ e^{\{n\}} = \frac{1}{2}\sum_l \beta_l(y_l^{\{n\}} - T_l^{\{n\}})^2 \ \ , \eqno(4) $$

\noindent where $T_l$ is the target output value for the $l^{th}$ output node.
The $\beta_l$ terms allow us to assign different weights to different
outputs, and thereby give more priority to determining certain outputs
correctly.

The most popular method for training the network is with the {\it
backpropagation algorithm} (Rumelhart, Hinton and Williams 1986a,b),
in which we determine the gradient of the error function with respect
to each of the weights in the network.  The network is first
initialized by setting the weights to small random values. We then
pass one of the training vectors (spectra) through the network and
evaluate the outputs, $y_l$, as well as all the intermediate node
values in the network (the vectors {\bf p} and {\bf q}).
Differentiating equation 4 with respect to a weight, $w_{k,l}$, in the
final weights layer and substituting for $y_l$ from equation 3 gives

$$ \frac{\partial e}{\partial w_{k,l}} =
\beta_l (y_l - T_l) q_k \ \ , \eqno(5) $$

\noindent 
where we have used the fact that the weights are causally independent
of one another and of the node values. (We have dropped the $n$
superscript to make things less crowded.) We can determine the
gradient of $e$ with respect to a weight, $w_{j,k}$, in the previous
weights layer in the same way, but now $q_k$ is a function of
$w_{j,k}$,

$$ \frac{\partial e}{\partial w_{j,k}} =
 \sum_l \beta_l (y_l - T_l) \frac{\partial y_l}{\partial w_{j,k}} = 
 \frac{\partial q_k}{\partial w_{j,k}} \sum_l \beta_l (y_l - T_l) w_{k,l}
\ \ . \eqno(6) $$

\noindent 
The remaining partial derivative can then be evaluated from equation
2.  Thus we see that the error gradient with respect to any weight in the
network can be evaluated by {\it propagating} the error dependency back
through the network to that weight. This can be continued for any
number of layers, up to the input layer, giving us the complete error
gradient vector, $\partial e/ \partial {\bf w}$, where ${\bf w}$ is
the set of all network weights.
 
Having evaluated the error gradient vector, there are a number of ways in
which it can be used to train the network. The most common is the {\it
gradient descent} process, in which we adjust the weight
vector in the direction of the negative of the gradient vector, i.e.\

$$ \Delta {\bf w} = - \mu \frac{\partial e}{\partial {\bf w}} \ \ . \eqno(7) $$

\noindent The factor $\mu$ determines how large a step is made, and typically
has a value between 0.1 and 1.0. The gradient is then recalculated using the new
values of the weights, and over many iterations the weights should
move towards a value which represents a small value of the error, $e$.
For training to strictly converge (i.e.\ reach an exact minimum),
$\mu$ would have to decrease to zero during training, although in
practice a sufficiently small error can often be achieved without this.

Of course, we are generally interested in getting the network to
generalize its input--output mapping for a range of types of objects
(e.g.\ different classes of stars), so we must correspondingly use a
range of objects which are representative of the problem. We therefore
apply the above training algorithm successively to each vector in a
training sample. We may either update the weights after each vector is
passed, or we can keep the weights fixed and update only after the
errors have been calculated for all vectors in the training set. In
this latter case, known as {\it batch} training, we update using the
the average error,

$$ E = \frac{1}{N} \sum_{n=1}^{n=N} e^{\{n\}} \ \ , \eqno(8) $$

\noindent
and the corresponding average error gradient vector.

The network training is a nonlinear minimization process in $W$
dimensions, where $W$ is the number of weights in the network.  As $W$
is typically large, this can lead to various complications.  One of
the most important is the problem of local minima. Clearly, training
using the gradient descent technique will stop upon reaching the
bottom of a minimum, but this may only be a {\it local} minimum, not
the {\it global} minimum, and may correspond to a much larger
error. To help avoid local minima, a {\it momentum term} is sometimes
added to the weight update equation 7, in which a fraction, $\nu$, of
the previous weight update is added to the current one. This provides
some ability to ``escape'' from small local minima: the larger is
$\nu$ the larger are the minima which can be ignored, although
possible at the expense of slower convergence.  A brute force approach
to avoiding local minima is to retrain the network several times from
different initial random weights, and see whether each case converges
to the same solution. If the majority do, this could be taken as the
best solution. Alternatively, we could use all of these networks
together in a {\it committee}, in which the outputs from the networks
are averaged.

A common difficulty is knowing when to stop training, as true
convergence in all but very simple (or fortuitous) cases is rarely
achieved. Without searching the entire weights space, it is generally
impossible to know whether the minimum reached is the global one or
just a local one, and if local, how much better the global minimum
is. Furthermore, the minimum may have a long, almost horizontal bottom
(thinking analogously in three dimensions), so it will often be
necessary to terminate training when the gradient falls below some
small value. Setting the convergence gradient too small will slow
convergence; too large and the search may terminate prematurely.

Fig.~3 shows several examples of how the error in equation 8 changes
during training. The dotted line shows the error on the actual
training data. However, we really want to test whether the network has
generalized its mapping, rather than ``memorized'' the specific
training data.  The solid line therefore shows the error evaluated on
a separate test data set, which is not used to modify the
weights. Under certain circumstances it is possible for the network to
fit the noise in the training set, in which case it is said to have
{\it overtrained}. This can be seen in Fig.~3c, where the error on the
test data starts to rise at some point, although the error on the
training data continues to fall. Frequently, both errors continue to
drop monotonically until levelling out (Fig.~3a), which is
desirable. (A difference in the errors finally achieved may reflect
differences in the training and test data sets.)  Fig.~3b shows an
example of where the error on the test data rises briefly, and then
drops: we should not, therefore, cease training at the first sign of a
rising test error. In certain cases the error can be very noisy
(Fig.~3d), oscillating around a minimum value, probably due to $\mu$
in equation 7 being held constant. Anecdotal evidence suggests that
this is more likely to occur when multiple outputs are used, although
it depends on the minimization algorithm used.

\noindent
\midinsert
\vskip 8.5cm
\includegraphics{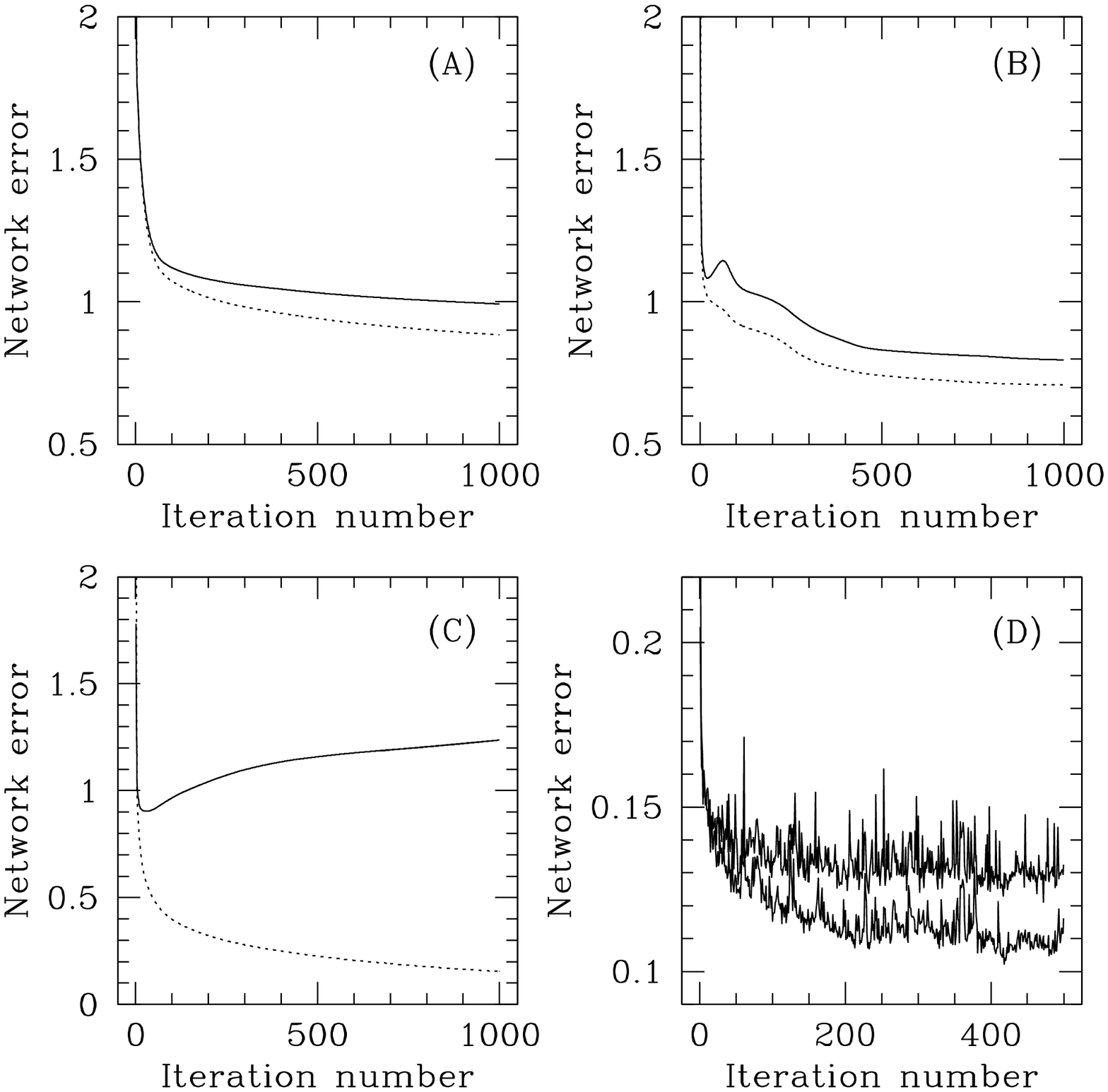}
\vskip 0.3cm
{\eightpoint   
\noindent
{\bf Figure 3.} Examples of the evolution of the network error during
training. The dotted (solid) line is the error evaluated on the
training (test) data (except in case D where the lower error is for
the training data).  (A),(B) and (C) are for networks with a single
continuous output. (D) is for a 50:5:5:3 network with probabilistic
outputs. All cases used gradient descent with constant $\mu$ and $\nu$
terms. See Bailer-Jones (1996) for more details.  }
\endinsert

Gradient descent only uses knowledge of the local gradient, which will
not, in general, point towards the global minimum, thus requiring many
steps to reach this minimum. Numerous other optimization techniques
exist to overcome this problem. One is the method of {\it conjugate
gradients}, in which a more efficient gradient is defined using the
second order derivatives (albeit implicitly) of the error with respect
to the weights. Several other methods make direct use of the second
derivatives (or {\it Hessian} matrix) to obtain more rapid
convergence.  We refer the interested reader to any number of books on
optimization techniques for further details.

The mapping produced by a neural network is an interpolation of the
training data. Whenever data are interpolated, there needs to be some
kind of complexity control to prevent overfitting the data. For
example, if we imagine fitting a curve through five noisy data points
in a two-dimensional space, then -- unless the data are co-linear -- a
fourth order polynomial model will give a better fit (a perfect one in
fact) than a straight line. This is unlikely to be appropriate,
though, as the data contain noise. Hence there is a trade off between
obtaining a good fit and not fitting the noise, and a compromise can
be achieved through the use of a {\it regularization} technique.  A
very basic approach is to split the training sample into two parts,
and during training on one half, monitor the total error on the other
half. As seen in Fig.~3, the test error {\it may} start to rise after
some number of iterations, indicating that the network has begun to
fit the noise (or {\it overfit} the data). Training could then be
stopped just before the error starts to rise. However, experience
shows that in some cases the training error never rises, even after a
very large number of iterations have occurred and a small error has
been achieved.

Another approach to regularization is {\it weight decay}. A high order
fit to data is characterized by large curvature of the mapping
function, which in turn corresponds to large weights. We can penalise
large weights by adding the following term to the training error in
equation 8

$$ \alpha \frac{1}{2} \sum_i w_i^2 \ \ , \eqno(9) $$

\noindent
where the sum is over all weights in the network and $\alpha$ is some
constant.  

Many neural network applications in astronomy have used gradient
descent for training, to the extent that some people think gradient
descent and neural networks to be synonymous.  This is not the case:
the feedforward neural network is a general means of representing a
nonlinear relationship between two domains, with the feature that the
error gradient is easily calculable in terms of the free parameters
(the weights) of the model. What optimization technique is then used
to determine the best solution for the weights is really a separate
issue. Gradient descent is often used because it is easy to code, but
it may not be the best choice.

\vskip 12 pt
\centerline{\bf 4.\ A Bayesian perspective}
\bigskip
\noindent
A rather different approach to interpolating data with a feedforward
network is provided within a Bayesian probabilistic framework.  If we
go back to first principles, then we see that the problem we are
trying to answer with the neural network is fundamentally
probabilistic: Given the set of training data, $D$, and some
background assumptions, $H$, what is the most probable output vector,
$\yv^*$, for some given input vector, $\xv$ (not in $D$)? When phrased
this way, we see that we are not really interested in the network
weights, and, moreover, an ``optimal'' set of network weights is not
actually required.  For example, if we have trained a network and
found a good mapping, we may retrain it and find a different set of
weights which give a mapping which is almost as good. Thus, while a given
input will yield much the same output as before, the weights may be
very different. In the optimization approach we ignore all but one
solution (except when using committees), whereas in a Bayesian
approach we combine all such solutions by the process of {\it
marginalization}. Specifically, if the output of the network with
weights $\wv$ and input $\xv$ is written $\yv(\xv, \wv)$, then

$$ \yv^* = \int \yv(\xv, \wv) P(\wv|D,H) d \wv  \ \ . \eqno(10) $$

\noindent
In other words, we weight each possible output by the probability,
$P(\wv|D,H)$, that the training data would produce those weights.
This {\it posterior} probability for the weights is given by Bayes'
Theorem

$$ P(\wv|D,H) \propto P(D|\wv,H) P(\wv|H) \ \ . \eqno(11) $$

\noindent
The first term on the right hand side of this equation is the {\it
likelihood}. It is the probability that a network with a given set of
weights produces the targets in the training data, D, when the inputs
from these data are fed in. If we assume a Gaussian error model for
the discrepancies between outputs and targets, this is just

$$ P(D|\wv,H) \propto \exp{-E} \ \ , \eqno(12) $$

\noindent
where $E$ is the sum of squares error from equations 8 and~4.  The
other term in equation 11 is the {\it prior probability} over the
weights, and represents our prior belief of what the weights should
be. In this Bayesian framework, the weight decay regularizer mentioned
earlier (equation 9) is naturally incorporated into this prior.

From equations 11 and 12 we can evaluate all of the terms under the
integral in equation 10. However, this integral must usually be
evaluated numerically with Monte Carlo methods. This involves
replacing the integral with a summation over $\yv(\xv, \wv)$ for a
large number of values of $\wv$ drawn from the distribution
$P(\wv|D,H)$. In this Bayesian approach, the optimization of the
weights is replaced with the integration over all possible weight
values, so no ``optimum'' weight vector is (or has to be)
calculated. While this approach is theoretically more satisfying than
an optimization, it is time consuming, as the marginalization must be
carried out for each new input vector we wish to classify. Partly for
this reason there exists an approximation to this approach which,
although still within the Bayesian framework, returns the problem to
one of optimization. There are several advantages to the Bayesian
approach, such as automatic complexity control (avoidance of
overfitting) and a natural way for determining the uncertainty in each
network output. More on the Bayesian approach for feedforward neural
networks can be found in (MacKay 1995).  For a general introduction to
Bayesian methods, Sivia (1996) is recommended.

\vskip 12 pt
\centerline{\bf 5.\ Implementation issues}
\bigskip
\noindent
Now that the basic principles of neural networks have been laid out,
it is necessary to ask how they are used in practice to solve a
problem.  We illustrate this with the problem of stellar spectral
classification.

\vskip 12 pt
\centerline{\bf 5.1.\ Training and testing data}
\bigskip
\noindent
The first thing which must be specified is the training data set,
consisting of spectra preclassified on the classification system of
interest. It must be emphasized that the network cannot be more
accurate than these preclassifications, so considerable effort must
often be spent in assembling the training data.  We must also ensure
that the examples in the training data set cover the full range of
types to which we will want to apply the trained network:  We cannot
expect a network trained only on B stars to classify M stars well, as
the characteristic features of M stars are not present in B star
spectra. Conversely, we do expect a network trained only on early and
late B stars to classify mid-B stars well, as B stars all have
qualitatively similar spectra.

Once a network has been trained, we want to evaluate its performance,
and this must be done using a separate data set. As mentioned earlier,
it is possible for a network to fit the noise and thus memorize the
specific training data set rather then generalize from it.  This can
be a particular problem if the data are noisy or if the network is
very complex (i.e.\ has many hidden nodes so is flexible to model many
nonlinearities). We can test for this by training the network on only
a subset of the available data and then evaluating it on the rest, ensuring
that the two data subsets have similar distributions of objects.

\vskip 12 pt
\centerline{\bf 5.2.\  Form of the inputs}
\bigskip
\noindent
The input to the network is some measured feature vector. In the case
of stellar classification this will be a spectrum or a list of fluxes
through various filters, or both. We could also input other
measurements we believe to be relevant, such as the parallax or
interstellar extinction. The inputs need not be contiguous or in any
particular order. A stellar spectrum typically contains a large amount
of redundant information (from the point of view of general
classification), so using the entire spectrum as the input may not be
necessary. If we can reduce the number of input nodes, we reduce the
number of weights and hence the dimensionality of the mapping
function. This results in faster training, and may reduce the chance
of getting stuck in local minima. It also increases the data/weights
ratio (see section 5.5) and density of the data, thus providing a more
reliable mapping. There are a number of approaches to {\it
dimensionality reduction}. The simplest is to remove any inputs which
we know, or suspect, to have little or no influence on the problem in
hand. A more sophisticated approach is with Principal Component
Analysis (PCA), which finds the most discriminatory linear
combinations of the inputs. By removing the less significant principal
components, we can compress the data with negligible loss of
discriminatory information, and filter out some noise too. PCA is
discussed in the article by Singh, Bailer-Jones \& Gupta in these
proceedings.

\vskip 12 pt
\centerline{\bf 5.3.\ Output modes}
\bigskip
\noindent
The output vector is the set of classifications.  For example, we may
have one output for MK spectral type (SpT) and another for luminosity
class (LC). We could alternatively use separate networks for each
parameter.  As the network outputs are real numbers, these classes are
represented on a continuous scale. Thus the network will be able to
produce intermediate classes. This is useful for the spectral type
parameter, as spectral types are simply points on what is really a
continuous scale (closely related to effective temperature). We refer
to this way of representing the outputs as {\it continuous
mode}. Luminosity classes, on the other hand, have more claim to being
discrete, for which we may want to use the network in {\it
probabilistic mode}, whereby each class is represented by a separate
output.  By confining each output to the range 0--1 (e.g.\ by replacing
equation 3 with the function $0.5 \times [1 + \tanh]$), we can
interpret the output as the probability that the input is of that
class.  For a five class problem, the target vector for an object in
class 2 would be $(0,1,0,0,0)$. When we apply the trained network to
new data, we would assign the class with the largest probability,
although if that probability is below some threshold, or not much
larger than the probability of some other class, then we may prefer to
label this object as uncertain.

Clearly, the probabilistic mode offers more information than the
continuous one. For example, if the input is a composite of two
classes (e.g.\ two stars of different brightnesses), we can represent
both in the target vector (e.g.\ 0,0.25,0,0.75,0). Thus if the network
thinks an input is composite, it can (in principle at least) tell us
this.  The only way we could model binary star spectra in
continuous mode is to have two spectral type outputs, although then
we also need some means of knowing when the input spectrum is just a
single star.  See Weaver (2000) for an attempt to classify binary
spectra with neural networks. A disadvantage of the probabilistic mode
is that it may result in many more network weights that a network in
continuous mode.

\vskip 12 pt
\centerline{\bf 5.4.\ How many hidden nodes and layers are required?}
\bigskip
\noindent
Deciding upon the appropriate number of hidden nodes and layers is
largely a matter of experience. With many problems, sufficient
accuracy can be obtained with one or two hidden layers and 5--10
hidden nodes in those layers. There is a theorem which states that a
network with one hidden layer can approximate any continuous function
to arbitrary accuracy, provided it has a sufficient number of hidden
nodes (Hornick, Stinchcombe and White 1990). In practice,
such a large number of nodes may be required that it is more efficient
to go to a second hidden layer. This can be seen in the work of
Bailer-Jones et al.\ (1998b). Fig.~4 shows how the error in
determining the spectral type (in continuous mode) varies as a
function of the number of hidden nodes in a network with just one
hidden layer.  The error is already reasonably small for a network
with just one hidden node, which is equivalent to a network with no
hidden layer. Adding a few hidden nodes improves the error, but once
there are 5--10 nodes, the error does not decrease further. However,
if we add a second hidden layer, then a network with 5 nodes in each
of these layers produces a $\sigma_{\rm 68}$ error of 0.85 SpT, a
statistically significant improvement. A similar improvement using a
second hidden layer was seen again in a related (but more complex)
problem (Bailer-Jones 2000).

\noindent
\midinsert
\vskip 6.5cm
\includegraphics{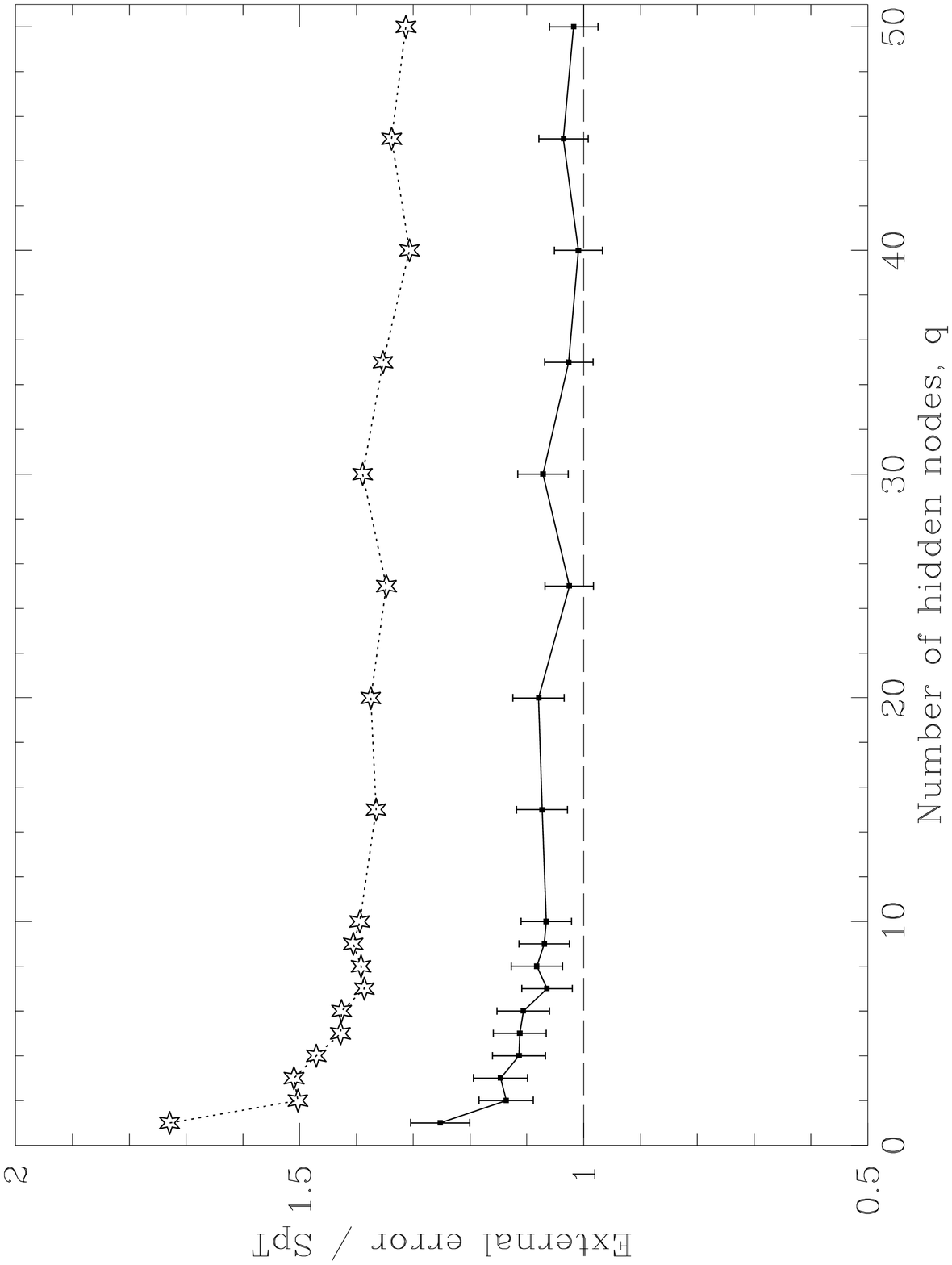}
\vskip 0.3cm
{\eightpoint   
\noindent
{\bf Figure 4.} Variation of the network classification error as a
function of the number of hidden nodes, $q$. The solid line is for
\sig68\ and the dotted line for \sigrms; they represent the core and
outliers of the error distribution respectively (see Bailer-Jones et
al.\ 1998b).  The network architecture is 50:$q$:1 with a continuous
spectral type output.  The error bars are $3 \times \delta$, where
$\delta$ is the standard deviation in \sig68\ due to a finite sample
size. The mean RMS error in the target classifications themselves is
0.65 SpT.  }
\endinsert

\vskip 12 pt
\centerline{\bf 5.5.\ Ensuring a well-determined network solution}
\bigskip
\noindent
An important consideration in setting up the network is the ratio of
data to weights. If we train a network with one output using $N$
training vectors, we have a total of $N$ error measurements with which
to determine the $W$ weights. Thus we have a system of $N$ equations
with $W$ unknowns, and, if the weights and training vectors are
independent of each other, we require $N > W$ to find a solution.
With fewer data the solution for the weights will be
underdetermined. Thus a 10:5:5:1 network (indicating 10 inputs, 5
nodes in each of two hidden layers, and one output) has 91 weights
(including the bias nodes), so will require at least as many training
examples. The data/weight ratio estimation is more complex with
multiple outputs (especially if they are correlated, as will be the
case in probabilistic mode), but if there are two independent outputs,
then the number of error measures is $2N$.

In practice, this ``overdetermination'' requirement may not be so
strict.  For example, Gulati et al.\ (1994) used only 55 training
examples to train a 161:64:64:55 probabilistic network. This network
has about 18,000 free parameters (each output is parametrized by about
15,000 weights), yet only of order 55 error measures were available to
determine them. In spite of this, a reasonably small error on a test
data set was obtained. We suspect that many weights were essentially
not trained, and, by staying at their small initial values, played no
role in the network. Additionally, correlations between the inputs
(flux bins of stellar spectra) could result in correlated weights in
the input--hidden layer. Both of these reduce the effective number of
free parameters. Other cases of apparently good results with formally
underdetermined weights appear in the literature, implying that
correlated or unused weights are not uncommon. Nonetheless, in many of
these cases simpler networks could probably have been used without
loss of accuracy.

\vskip 12 pt
\centerline{\bf 6.\ Example Applications}
\bigskip
\noindent
The paper by Gupta et al.\ in these proceedings gives an application
of a 93:16:16:17 network to classify 2000 bright sources from IRAS LRS
spectra.

Fig.~5 shows the performance of a committee of ten 50:5:1 networks
applied to the spectral classification problem, trained using gradient
descent. It shows the correlation between the network classifications
and the ``true'' classifications (those from the training catalogue),
and the histogram of the error residuals, for a test data set. There
is an even scatter about the ideal classifications in the left hand
plot, except for the hottest (early type) stars, where the network
assigns cooler classes.

\noindent
\midinsert
\vskip 5.8cm
\includegraphics{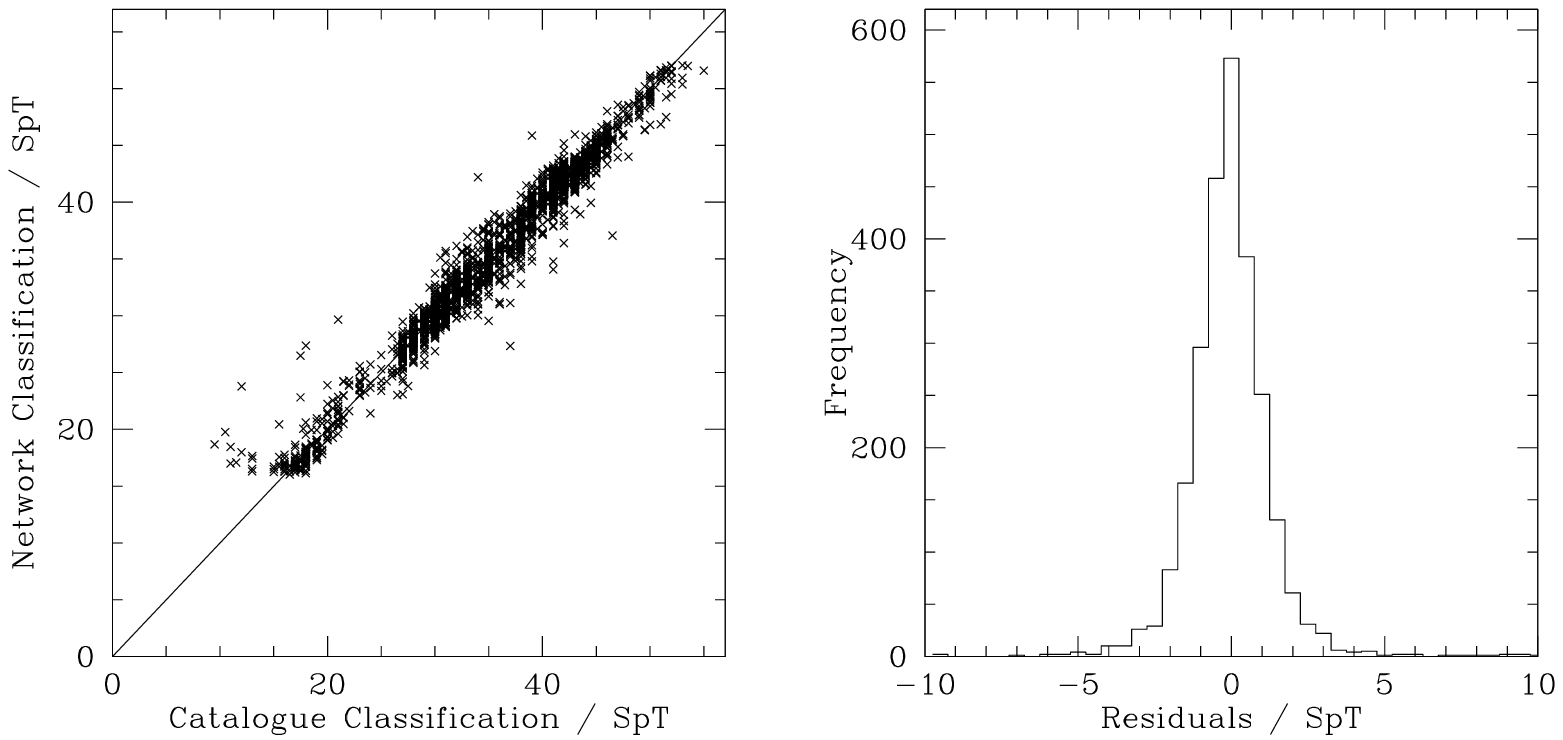}
\vskip 0.3cm
{\eightpoint 
\noindent
{\bf Figure 5.} Results from a committee of ten 50:5:1 networks with a
single continuous spectral type output. The left plot shows the
correlation between network and catalogue classifications (perfect
classifications would fall on the diagonal line), and the residuals
from this are plotted on the right.  The spectral types O3 to M9 have
been coded by the integers 1 to 57. The summary results are
$\sigma_{68} = 1.03$ SpT and $\sigma_{RMS} = 1.35$ SpT. The internal
error of the networks is 0.41 SpT. This is a measure of the average
differences in classifications which members of the committee assign
to a given spectrum.  }
\endinsert

A similar plot to that in Fig.~5 is not very useful for a network used
in probabilistic mode when there are only a handful of classes. In
such cases it is more useful to look at the {\it confusion matrix}.
Examples are shown in Fig.~6 for networks with three outputs,
corresponding to the three luminosity classes III, IV and V. In
probabilistic mode, we can also examine the confidence with which the
network classifies objects. An example is shown in Fig.~7, where we
plot the cumulative fraction of objects classified with a probability
greater than some amount.  This plot is for the 50:5:5:3 network with
the line+continuum spectra from Fig.~6.  Thus while 98.1\% of the
class V objects were correctly classified (i.e.\ a fraction 0.981 were
classified with a probability, $p>1/3$), only 77.5\% have a
probability of over 0.90. Part of the reason for this is that a
network with sigmoidal outputs can never produce an output of exactly
0.0 or 1.0, as that requires very large weights, something which
regularization tries to avoid. (See Bailer-Jones 1996 and Bailer-Jones
et al.\ 1998b for more details on these examples.)

\noindent
\midinsert
\vskip 8.5cm
\includegraphics{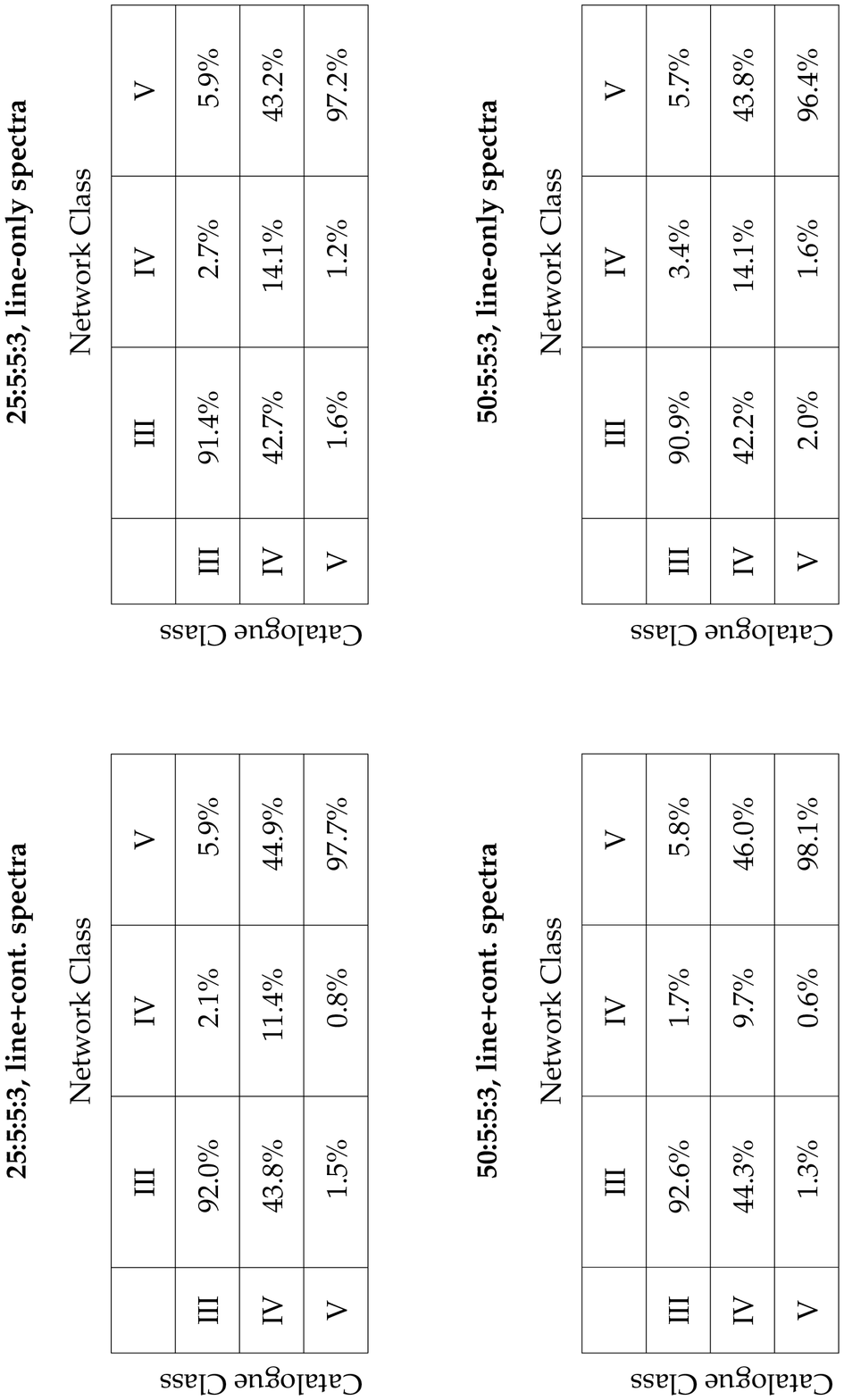}
\vskip 0.3cm
{\eightpoint   
\noindent
{\bf Figure 6.}  Confusion matrices for four different network models
differing in the number of PCA inputs and whether or not the continuum
has been subtracted from the spectra.  Each matrix lists the
percentage of spectra which have been classified correctly and
incorrectly: in the top-left matrix the network classifies 1.5\% of
spectra which are class V in the catalogue as class III, whereas it
correctly classifies 97.7\% of class Vs. In all four cases the
networks' do not believe in the existence of luminosity class IV
(random assignments would lead to about 33\% in the middle row).  }
\endinsert

\noindent
\midinsert
\vskip 13.5cm
\includegraphics{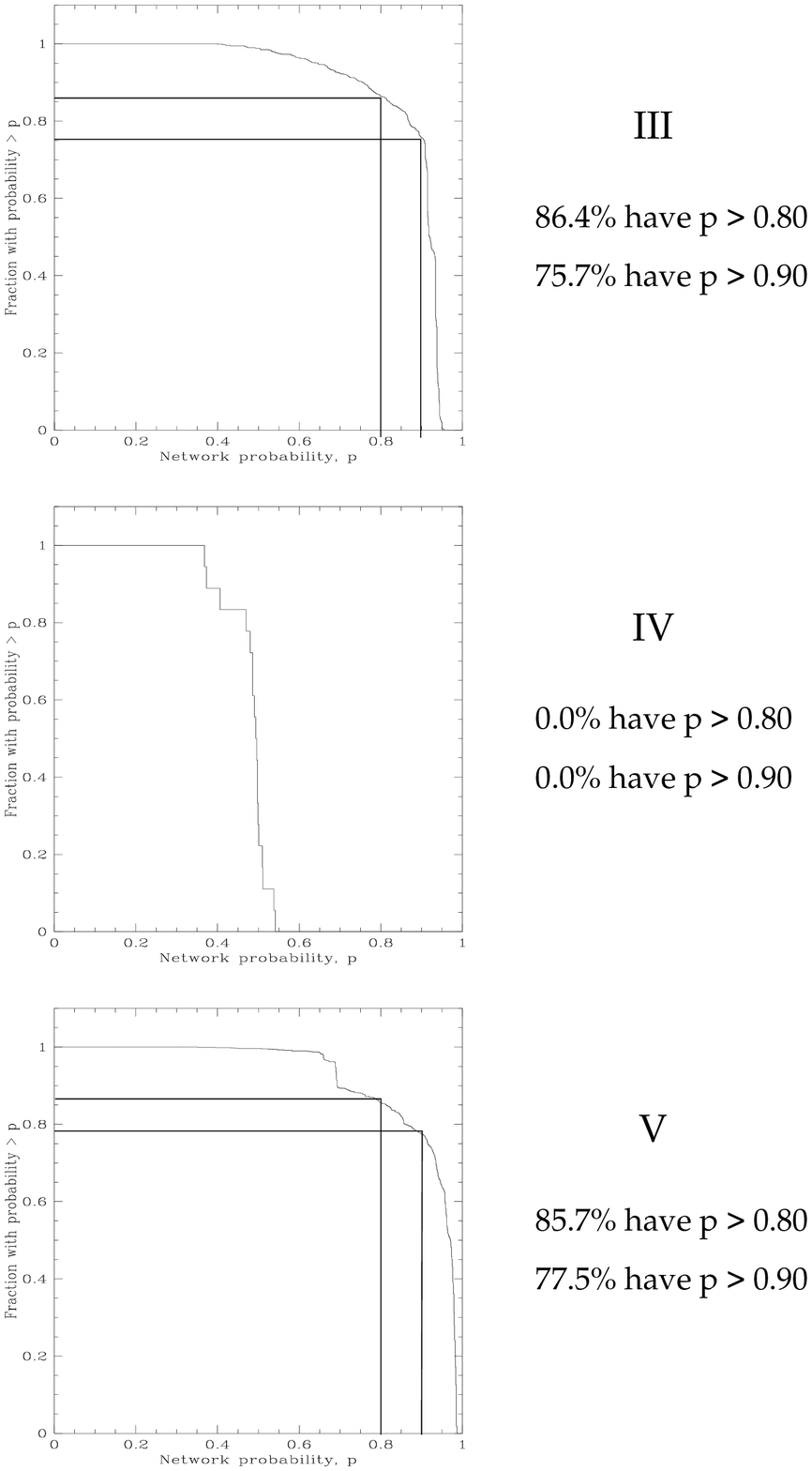}
\vskip 0.3cm
{\eightpoint   
\noindent
{\bf Figure 7.}  Confidence of correct network classifications based
on network output probability.  Each figure is a cumulative plot of
the fraction of spectra which have been classified with a probability
$>p$, for the luminosity classes III, IV and V.}
\endinsert

\vskip 12 pt
\centerline{\bf 7.\ Understanding the network}
\bigskip
\noindent
The underlying principle of a neural network is that it learns its
input--output mapping from the relationship intrinsically present in a
preclassified data set. Unlike rule-based classification algorithms,
the networks are not explicitly told anything about the relevance of
certain inputs in determining the output. This relevance information
is expressed in the optimized weight vectors, and in principle we can
look at the values of the weights to find out which inputs are
significant in determining the outputs.  In practice, however, this
may be rather difficult on account of there being more than one hidden
node (which is necessary to get a nonlinear mapping). For example,
weights leading from a relevant input to different nodes in the hidden
layer could be both small and large in magnitude. There is no
constraint on the network training to provide easily-interpretable
weights, although such a one may be possible. Furthermore, there may
be many minima in the weights space which have approximately
equal errors, and the corresponding ``optimal'' weight vectors may
give conflicting opinions about the relevance of the inputs. A
degeneracy in the weights is made more likely if there are redundant,
or correlated, inputs, which is often the case with stellar
spectra. Ultimately, obtaining both an accurate model of a complex
problem {\it and} a good understanding are conflicting demands, as
understanding often requires simplification.

\noindent
\midinsert
\vskip 17.0cm
\includegraphics{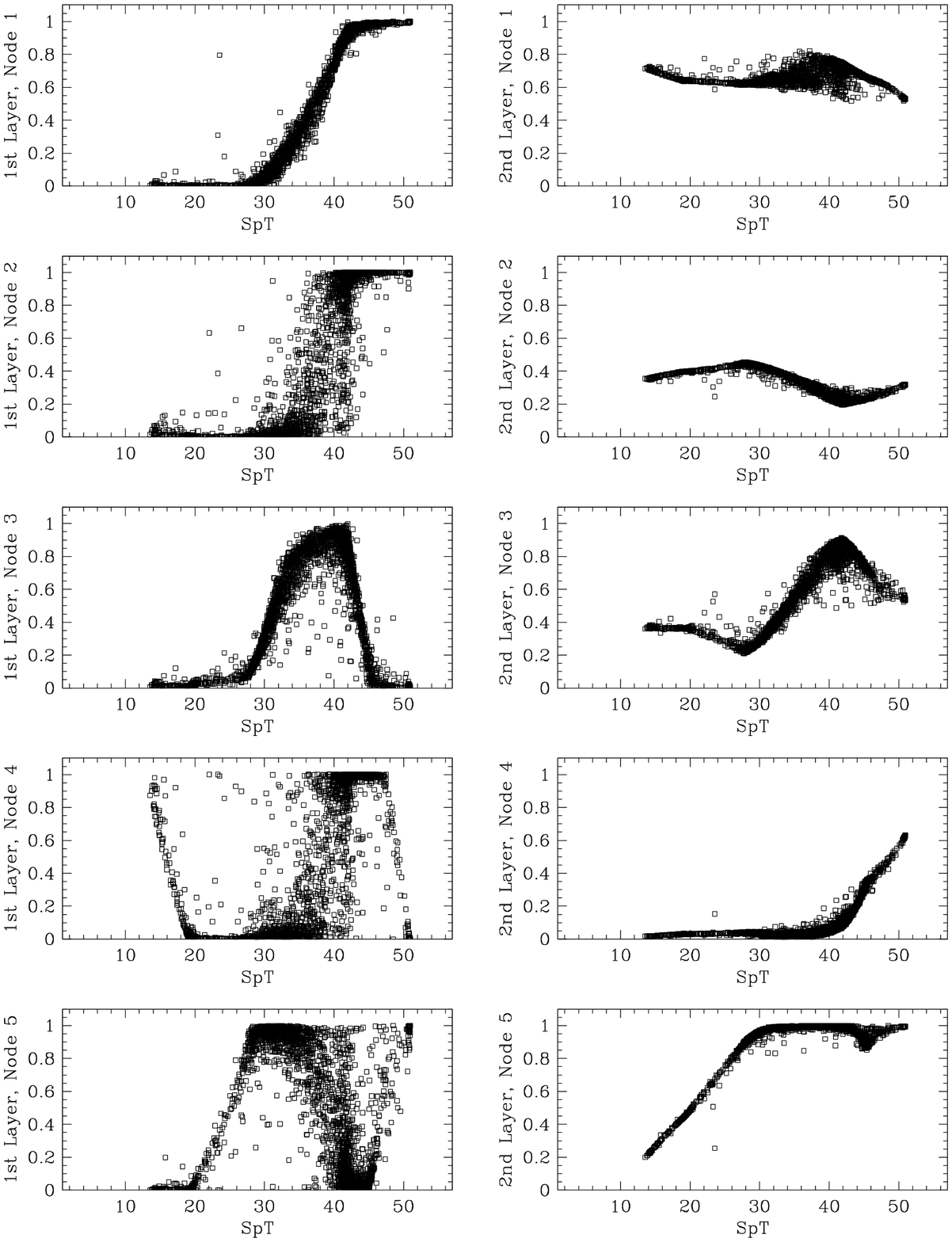}
\vskip 0.3cm
{\eightpoint 
\noindent
{\bf Figure 8.}  Correlation between the node values and outputs for a
network trained to give spectral types in continuous mode.  The column
on the left (right) represents the first (second) hidden layer of
nodes in a 50:5:5:1 network.  Each point is the node value produced by
one input vector in an test data set. The transfer function in each
layer (including the subsequent output layer) is a sigmoid function,
which restricts the node values to the range 0--1.  }
\endinsert

We can nonetheless get some idea of {\it how} a network produces its
classifications. Fig.~8 shows the correlations between the values on
the hidden nodes of a network and the outputs from that
network. Looking at the first hidden layer, we see that no single node
shows a low-scatter, monotonic relationship with spectral type. If one
did, then that alone could be used for spectral type classification
and we could dispense with the network altogether. Generally, the
correlations are nonlinear, and with large scatter.  However, node 1
does show linearity over a limited part of the spectral range (SpT =
28--42, or F0--K0). This in interesting, as it means that for classes
F and G this node could serve as the output node and we would get
reasonable classifications. Looking back at equation 1, we see that if
we were to re-parametrize the spectral type as $tanh^{-1}$ SpT, we
could classify F and G stars using a simple linear regression on the
spectral indices (as $tanh^{-1}$ is monotonic over the range of
interest).  Nodes 3--5 also show linear correlations over small
portions of the spectral type range. However, all nodes show either
some insensitivity (due to saturation of the node values) or large
scatter in the correlation, for at least some part of the spectral
type range.

When these values are passed to the second hidden layer, the situation
is somewhat different. The correlations are still nonlinear, but there
is much less scatter. Furthermore, node 5 has a low scatter, {\it
linear} correlation for spectral types earlier than about F3 (SpT =
30), and node 4 shows a linear correlation for types later than about
K0 (SpT = 42). Node 2 (or 3) alone could reasonably classify all intermediate
types. Thus we see that there is some localization in the network,
with certain nodes taking on the job of identifying certain ranges of
the target variable. There is, of course, no significance in the
numbering of the nodes within a layer, as the nodes could be shuffled
with no change in the final output.

\vskip 12 pt
\centerline{\bf 8.\ Summary}
\bigskip
\noindent
We have introduced the basic concepts of feedforward neural networks.
Although the type introduced, the multilayer perceptron, is only one
of many types, it is very important as judged by its wide and numerous
applications.  We have described the backpropagation algorithm for
training the network, as well as some of the ways in which it can be
implemented to optimize the weights. We also briefly looked at a
Bayesian approach to modelling with a network, which marginalizes over
the weights rather than optimizing them. A number of implementation
issues were discussed and example applications were presented.

Although it is interesting to see how these models were originally
inspired by attempts to model brain behaviour, these artificial neural
networks really have very little in common with biological neural
networks, and, in our opinion, are better understood in purely
mathematical terms. Whilst they may be powerful models, there is
nothing ``mysterious'' about them. They are simply parametrized
nonlinear models: the input--output function of a two-hidden layer
model is described entirely by equations 1--3. They are no more a
``black box'' than is a straight line fit to two-dimensional data. The
nonlinearity and multidimensionality of neural networks are present
because real-world problems are often complex. If we have difficulty
understanding how they work, then it is because an {\it accurate}
description of real-word problems often requires that we avoid the
simplifying assumptions which typically make easily-understood
descriptions only approximate.

We finish by stressing that the purpose of these models is to
generalize the relationship between a data domain and a parameter
domain which is intrinsically present in a preclassified data set.
They are useful because they can model complex relationships in a
large number of parameters, and generalize this to areas of the input
space where training data are sparse. This has potential applications
to large, multiparameter astronomical surveys (see Bailer-Jones, these
proceedings). But the models can only be as good as the data on which
they are trained (although they can be a lot worse if trained badly),
so construction of an accurate and relevant training data set is of
fundamental importance.

\bigskip
\centerline{\bf References}
\bigskip
{\eightpoint\parindent=0pt\everypar={\hangindent=0.5 cm}

Bailer-Jones, C.A.L., 1996, PhD thesis, Univ.\ Cambridge

Bailer-Jones, C.A.L., 2000, A\&A, 357, 197

Bailer-Jones, C.A.L., MacKay, D.J.C., Withers, P.J., 1998a, Network:
Computation in Neural Systems, 9, 531

Bailer-Jones, C.A.L., Irwin, M., von Hippel T., 1998b, MNRAS, 298, 361

Beale, R., Jackson, T., {\it Neural Computing: An Introduction}, 1990,
Institute of Physics: Bristol

Bishop, C.M., {\it Neural Networks for Pattern Recognition}, 1995,
Oxford University Press: Oxford

Gulati, R.K., Gupta, R., Gothoskar, P., Khobragade, S., 1994, ApJ, 426, 340

Hertz, J., Krogh, A., Palmer, R.G., 1991, {\it Introduction to the
Theory of Neural Computation}, Addison-Wesley: Wokingham, England

Hornick, K., Stinchcombe, M., White, H., 1990, Neural Networks, 3, 551

MacKay, D.J.C., 1995, Network: Computation in Neural Systems, 6, 469

Rumelhart, D.E., Hinton, G.E., Williams, R.J., 1986a, in 
D.E.\ Rumelhart, J.L.\ McClelland J.L, the PDP Research Group, eds, 
Parallel Distributed Processing: Explorations in the
  Microstructure of Cognition, MIT Press: Boston, p.\ 318

Rumelhart, D.E., Hinton, G.E., Williams, R.J., 1986b, Nature, 323, 533

Sivia, D.S., 1996, {\it Data Analysis. A Bayesian Tutorial}, Oxford
University Press: Oxford

Weaver, W.B., 2000, ApJ, 541, 298

}                                         
\end